\documentclass[aps, prl]{revtex4-2}  % chktex 8
\usepackage[utf8]{inputenc}
\usepackage[english]{babel}
\usepackage{amsmath,amsfonts,amssymb}
\usepackage{graphicx}
\usepackage{float}
\renewcommand{\vec}{\mathbf}
\newcommand{\sign}[1]{\text{sign} (#1)}

\begin{document}
%TC:ignore
\title{Incompressible Energy Spectrum from Wave Turbulence}
\author{Marcos A. G. dos Santos Filho}
\affiliation{Department of Physics, Federal University of São Carlos, 13565-905, São Carlos, SP, Brazil}
\author{Francisco E. A. dos Santos}
\affiliation{Department of Physics, Federal University of São Carlos, 13565-905, São Carlos, SP, Brazil}
\date{\today}
%TC:endignore
%TC:break Abstract
\begin{abstract}
    Bose--Einstein condensates with their superfluidity property provide an interesting parallel to classical fluids.
    Due to the Kolmogorov spectrum of homogeneous turbulence the statistics of the incompressible velocity field is of great interest, but in superfluids obtaining quantities such as the statistics of the velocity field from the macroscopic wavefunction turns out be a complicated task; therefore, most of the work up to now has been numerical in nature.
    We made use of the Weak Wave Turbulence (WWT) theory, which provides the statistics of the macroscopic wavefunction, to obtain the statistics of the velocity field, which allowed us
    to produce a semi analytical procedure for extracting the incompressible energy spectrum in the WWT regime.
    This is done by introducing an auxiliary wavefunction that preserves the relevant statistical and hydrodynamical properties of the condensate but with a homogeneous density thus allowing for a simpler description of the velocity field.
\end{abstract}
\maketitle

%\linenumbers
\section{Introduction}

Bose--Einstein condensates were initially discovered as a consequence of Einstein applying the statistical description proposed by Bose to ideal non-interacting gas of \emph{quanta}.
In Einstein's observations of quantum ideal , it was noted that the Bose statistics allowed for a macroscopic occupation of the system ground state bellow a certain \emph{critical} temperature.
The macroscopic occupation of a single energy state noted by Einstein is known as a Bose--Einstein condensate (BEC) and manifests both in the superfluid phase of $^{4}\text{He}$ and in ultracold dilute , among other systems.

Dilute gasses have much lower volumetric density when compared with other types of matter such as liquids, solids or air.
In order to observe quantum phenomena like Bose--Einstein condensation much lower temperatures are needed when compared to liquid helium however, due to the weak interatomic interaction of dilute gasses, much larger condensate fractions can be obtained.
The experimental realization of BEC in dilute gasses was achieved through the combination of laser and evaporative cooling techniques as well as advances in trapping arrangements, leading to two separate groups managing to cool a cloud of atoms below the necessary critical temperatures and to observe a macroscopic occupation of the system's ground state in 1995~\citep{andersonObservationBoseEinsteinCondensation1995,davisBoseEinsteinCondensationGas1995}.
A detailed description of the properties of dilute gasses in the context of Bose--Einstein condensation can be found in Ref~\citep{pethick_bose-einstein_2008,griffin_bose-einstein_1995-1}.

The study of turbulence in a Bose--Einstein condensate (BEC) is typically approached using the Gross--Pitaevskii equation (GPE)~\citep{tsatsos_quantum_2016-1, barenghiPrimerQuantumFluids2016,dalfovoTheoryBoseEinsteinCondensation1999,madeiraQuantumTurbulenceBose2020},
\begin{equation}\label{eq:gpe}
    i\hbar\frac{\partial{\psi(\vec{r},t)}}{\partial{t}} = -\left[\frac{\hbar^{2}}{2m}\nabla^{2} + V(\vec{r}) + g|\psi(\vec{r}, t)|^{2}\right]\psi(\vec{r}, t),
\end{equation}
where $\psi$ is the condensate wavefunction, $g$ the interaction parameter, $m$ the system mass, and $V$ the trapping potential.
%In the following we will be discussing the case of a uniform potential taken to be $V=0$.
The GPE is particularly appropriate to describe systems where the interaction between the atoms is relatively weak.
This is usually the case of the so-called atomic Bose--Einstein condensates which are achieved in experiments with dilute gasses.

\subsection{Turbulence}
Turbulence in general is most often studied by exploring the incompressible Navier--Stokes equations,
\begin{equation}\label{eq:navierstokes}
    \begin{aligned}
        \frac{\partial{\vec{v}}}{\partial{t}} + \vec{v}\cdot{\vec{\nabla}{\vec{v}}} &= - \frac{1}{\rho}\vec{\nabla}{P} + \nu\nabla^{2}\vec{v},\\
        \vec{\nabla}\cdot\vec{v} &= 0,
    \end{aligned}
\end{equation}
where $\vec{v}$ is the fluid velocity, $\rho$ the fluid density, $P$ the fluid pressure and $\nu$ the fluid kinematic viscosity.
The bottom equation in~\eqref{eq:navierstokes} states that conservation of mass in incompressible fluids implies that $\vec{v}$ is a solenoidal field.
Generally, turbulence happens when the non-linear term, $\vec{v}\cdot\vec{\nabla}\vec{v}$, in the left-hand side of~\eqref{eq:navierstokes} is much larger than the last term in the right-hand side, $\nu{\nabla}^{2}\vec{v}$, which implies a system with relatively high velocity and low kinematic viscosity.
A quick indicator of these characteristics is the Reynolds number, $Re = \frac{vD}{\nu}$, where $D$ is a characteristic length dependent on the particular system, e.g., the diameter of a pipe, the length of a container, etc\ldots
Therefore, a large Reynolds number can serve as a qualitative indicator of the development of turbulence in the system.

\begin{figure}[ht]
    \centering
    \includegraphics[width=8.6cm]{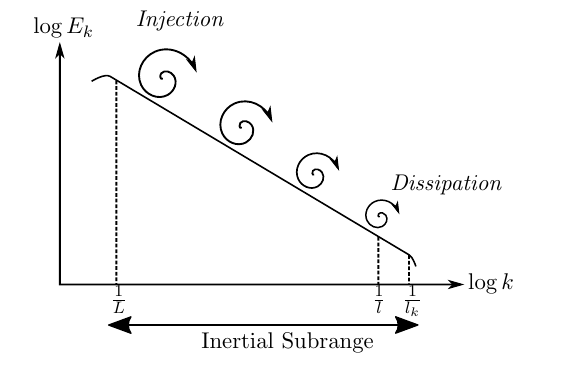}
    \caption{\label{fig:kolmogorovcascade}  Illustration of the energy cascade in the inertial range. Large eddies manifest at the energy injection scale and small eddies dissipate near the Kolmogorov length, $l_{k}$, scale.}
\end{figure}

The general understanding is that there is no complete theory of turbulence, largely due to the high complexity of solving coupled differential equations with chaotic behaviour, but even so, there has been a lot of effort in trying to understand turbulent flow.
In 1928 Richardson proposed~\citep{richardson_atmospheric_1926} that a well-developed turbulent system could be understood in terms of large eddies breaking down into smaller eddies which then in turn further break down in a self-similar process until the smallest eddies would dissipate due to the internal kinematic viscosity of the fluid.
An important quantity to understand the mechanism of the Richardson's cascade is the energy density per unit mass~\citep{leslieDevelopmentsTheoryTurbulence1973,mccombPhysicsFluidTurbulence2003} given, in terms of the statistics of the velocity field, by:
\begin{equation}
    E = \frac{1}{2}\langle{\vec{v}(\vec{r})\vec{v}(\vec{r})}\rangle  = \int\limits_{0}^{\infty}\mathcal{E}_{k}dk
\end{equation}
where $\mathcal{E}_{k}$ is the energy spectrum.
Kolmogorov proposed~\citep{kolmogorovLocalStructureTurbulence1991,frisch_turbulence:_1995} that for a homogeneous, isotropic and stationary turbulent flow, the energy per unit mass is distributed through a cascade process between the length scales (see Fig.~\ref{fig:kolmogorovcascade}) given by the characteristic length and the length of the smallest eddies before dissipation, also called the Kolmogorov length.
Inside this so-called inertial range, the energy spectrum under Kolmogorov's assumptions is:
\begin{equation}
    \mathcal{E}_{k} = K_{0}\varepsilon^{\frac{2}{3}}k^{-\frac{5}{3}},
    \label{eq:kolmogorovspectrum}
\end{equation}
where $\varepsilon$ is the energy injection rate and $K_{0}$ is an experimentally determined constant.
This result became known as the Kolmogorov spectrum and has been observed in a variety of fluid related phenomena.

\subsection{Quantum Turbulence}
The superfluid nature of quantum fluids like atomic BEC poses an interesting avenue for the exploration of turbulent flows.

The GPE, Eq.~\eqref{eq:gpe}, can be used to study both the dynamics of the condensate wavefunction and its hydrodynamical properties, the same approach can be applied in a qualitative manner to superfluid helium~\citep{pethick_bose-einstein_2008,griffin_bose-einstein_1995-1,dalfovoTheoryBoseEinsteinCondensation1999}.
From the GPE, we can obtain the continuity equation:
\begin{equation}
    \frac{\partial{\rho}}{\partial{t}} + \vec{\nabla}\cdot{\vec{j}} = 0
\end{equation}
where $\rho(\vec{r}) = |\psi(\vec{r})|^{2}$ is the particle density and $ \vec{j} = \frac{\hbar}{2\text{i}}(\psi^{*}\vec{\nabla}{\psi} - \psi\vec{\nabla}{\psi^{*}}) $ the particle current.
The condensate velocity field can then be defined by:
\begin{equation}\label{eq:velfield}
    \vec{v} = \frac{\vec{j}}{\rho{m}} = \frac{\hbar}{2m\text{i}}\frac{\psi^{*}\vec{\nabla}{\psi} - \psi\vec{\nabla}{\psi^{*}}}{|\psi|^{2}}.
\end{equation}

According to the Helmholtz theorem the velocity field can be separated into its incompressible, $\vec{\nabla}\cdot\vec{v}^{\perp} = 0$, and compressible, $\vec{\nabla}\times\vec{v}^{\parallel} = 0$, parts.
In analogy with classical turbulence we can write the incompressible kinetic energy per unit mass as:
\begin{equation}
    E = \frac{1}{2}\langle{\vec{v}^{\perp}(\vec{r}){\vec{v}^{\perp}}^{*}(\vec{r})}\rangle  = \int\mathcal{E}_{k}dk
\end{equation}
where it was assumed that the system is homogeneous and isotropic.
We can then, in principle, obtain the spectrum using the Fourier transform of the velocity field given by:
\begin{equation}
    \tilde{\vec{v}}^{\perp}(\vec{k}) = \frac{1}{{(2\pi)}^{3}}\int\vec{v}^{\perp}(\vec{r})e^{-i\vec{k}\cdot\vec{r}}dr^{3},
    %\tilde{\vec{v}}^{\perp{*}}(\vec{k}) & = \frac{1}{{(2\pi)}^{3}}\int\vec{v}^{\perp{*}}(\vec{r})e^{i\vec{k}\cdot\vec{r}}dr^{3}.
    \label{eq:enspectrum}
\end{equation}
As one can see, due to the nature of the velocity field in Eq.~\eqref{eq:velfield}, this integral can prove to be quite difficult to solve analytically mainly due to the $|\psi|^{2}$ term in the denominator of $\vec{v}$ which would cause discontinuities in any region where the particle density vanishes.

There has been a number of numerical simulations, of the GPE as well as other methods, to analyse the energy spectrum~\citep{kobayashiKolmogorovSpectrumSuperfluid2005,noreKolmogorovTurbulenceLowTemperature1997,bradleyEnergySpectraVortex2012,cidrimVinenTurbulenceDecay2017}.
Two regimes have been observed, the quasiclassical regime with $\mathcal{E}_{k} \sim k^{-\frac{5}{3}}$,
compatible with the Kolmogorov spectrum observed in classical turbulence~\citep{tsubotaQuantumTurbulence2008}, and the ultraquantum regime observed using the vortex filament model by ~\citet{baggaleyQuasiclassicalUltraquantumDecay2012}, with $ \mathcal{E}_{k} \sim k^{-1}, $ which was predicted to be compatible with Vinen turbulence.

\subsection{Wave Turbulence}
An important aspect of turbulence in BECs depends on the magnitude of the non-linear interaction term present in the Gross--Pitaevskii equation.
In the regime where the kinetic energy is dominant over the interaction energy, called the weak interaction regime, an analytical approach called Weak Wave Turbulence (WWT) theory~\cite{zakharovKolmogorovSpectraTurbulence1992,nazarenkoWaveTurbulence2011,nazarenkoWaveTurbulence2015,kolmakovWaveTurbulenceQuantum2014} can be applied.

WWT provides a way to obtain an elegant description of the waves interactions using a statistical approach.
To that end the Random Phase and Amplitude (RPA) hypothesis is assumed for the wavefunction in the momentum basis.
RPA assumes that both the phase and the amplitudes of the system waves are independent random variables and the phase is uniformly distributed~\cite{choiJointStatisticsAmplitudes2005}.
Such assumptions allow for a closure relation for the wave spectrum, $ n(k) \equiv \left\langle{\tilde{\psi}(\vec{k})\tilde{\psi}{(\vec{k}')}^{*}}\right\rangle\delta(\vec{k} - \vec{k}').  $
Also, a Wick contraction rule in which higher order correlators can be reduced to simple two component correlators can be used, e.g.:
\begin{equation}
        \left\langle{\tilde{\psi}(\vec{k}_{1})\tilde{\psi}(\vec{k}_{2})\tilde{\psi}^{*}(\vec{k}_{3})\tilde{\psi}^{*}(\vec{k}_{4})}\right\rangle = n(\vec{k}_{1})n(\vec{k}_{2})[\delta(\vec{k}_{3} - \vec{k}_{1})\delta(\vec{k}_{4} - \vec{k}_{2}) + \delta(\vec{k}_{4} - \vec{k}_{1})\delta(\vec{k}_{3} - \vec{k}_{2})].
\end{equation}

The evolution of the wave spectrum is given by the kinetic equation,
\begin{equation}
        \dot{n}_{k} = 4\pi\int\delta(\vec{k} + \vec{k}_{3} - \vec{k}_{1} - \vec{k}_{2})\delta(k^{2} + k^{2}_{3} - k^{2}_{1} - k^{2}_{2})n_{k_{1}}n_{k_{2}}n_{k_{3}}\left[\frac{1}{n_{k_{1}}} + \frac{1}{n_{k_{3}}} - \frac{1}{n_{k_{1}}} - \frac{1}{n_{k_{2}}}\right] dk_{1}^{3}dk_{2}^{3}dk_{3}^{3},
    \label{eq:kineticequation}
\end{equation}
where we took $\hbar=m=g=1$ for the sake of simplicity.
For a detailed description of the kinetic equation properties and applications one can turn to Ref.~\cite{zakharovKolmogorovSpectraTurbulence1992,nazarenkoWaveTurbulence2011}.

An indicator of sustained turbulence is the transference of energy freely between scales, for that we look for stationary solutions of equation~\eqref{eq:kineticequation}.
There are two types of stationary solutions for~\eqref{eq:kineticequation}, thermal equilibrium solutions which are given by the Rayleigh--Jeans spectrum,
\begin{equation}
    n_{k} = \frac{T}{k^{2} + \mu},
\end{equation}
with $T$ and $\mu$ being the temperature and chemical potential respectively~\cite{connaughton2005}, and two cascading solutions each relating to constant flux of the conserved quantities of the system, energy and number of particles.
Zakharov developed a formal approach to obtain these solutions, the Zakharov transformations, but in some cases the solutions can also be inferred through dimensional analysis as it is shown in Ref.~\cite{nazarenkoWaveTurbulence2011, nazarenkoWaveTurbulence2015}.

Due to conservation of energy, $E = \int{k^{2}n_{k}dk^{D}}$, we have the direct energy cascade, $n^{E}_{k} \sim k^{-D}$, while due to conservation of particles, $N = \int{n_{k}dk^{D}}$, we have the inverse waveaction cascade, $n^{N}_{k} \sim k^{-D+\frac{2}{3}}$, with $D$ representing the dimension of the system.
It should be mentioned that the energy cascade $n^{E}_{k}$ is log-divergent in the infrared limit, this nonlocality is usually log-corrected~\citep[see][chap.~15]{nazarenkoWaveTurbulence2011}.

The energy and particle transference through such cascades lead to evaporation of high energy particles and accumulation of low energy particles in the ground state, thus resulting in a non-equilibrium condensation process~\citep{nazarenkoWaveTurbulence2011}.
When the background condensate is large enough the interaction energy no longer can be considered weak and the 4-wave process described by the kinetic equation~\eqref{eq:kineticequation} is no longer valid.
The system can then be modelled by a background condensate of constant density plus a linear inhomogeneous perturbation, the WWT can also be applied to such a situation leading to an equivalent kinetic equation for a 3-wave system~\citep{fujimoto_bogoliubov-wave_2015,nazarenkoWaveTurbulence2011}.
In the following we will be focusing only on the three-dimensional case of the 4-wave regime.

There has been considerable effort to verify the WWT predictions numerically~\cite{proment_sustained_2012} as well as experimentally, both with harmonic~\cite{hennEmergenceTurbulenceOscillating2009,thompsonEvidencePowerLaw2014} and box like~\cite{navon_emergence_2016,navonSyntheticDissipationCascade2018} traps.
A comprehensive numeric investigation was conducted for the 3D case of WWT by \citet{proment_sustained_2012}, in this work a variety of situations were explored, with different methods of energy injection and dissipation.
Aside from the WWT cascades the authors of~\citep{proment_sustained_2012} also looked at the compressible and incompressible energy spectrum for some situations, this is of great interest since vortices were observed to be present in the 4-wave regime and in the transition to 3-wave regime most vortices should vanish, so that the compressible energy dominates the system.

Looking at the energy in the WWT, $E = \int{k^{2}n_{k}dk^{D}}$, we can predict the total energy spectrum, $\mathcal{E}(k)_{WT}^{(1D)} = k^{4+\alpha}$~\citep{promentEnergyCascadesSpectra2009}.
But it will only correspond with the incompressible kinetic energy spectrum when the conditions are such that the incompressible part of the fluid motion is largely dominant over the compressible part.
In the next section we will present a method to obtain the incompressible energy spectrum generally, as long as the kinetic equation is valid and the stationary solutions are known.

\section{Energy Spectrum in Wave Turbulence Regime}

Since the wave turbulence regime is well understood in a statistical sense, due to WWT, we will show that it is possible to obtain a semi-analytical  description of the incompressible energy spectra shown in equation~\eqref{eq:enspectrum}.

In order to study vortices and other structures in condensates one usually applies the Madelung transformation~\cite{pethick_bose-einstein_2008,davisBoseEinsteinCondensationGas1995}, $ \psi = \sqrt{\rho}e^{iS} $, where $S$ represents the phase field.
Applying this transformation to the velocity field~\eqref{eq:velfield} we obtain the following:
\begin{equation}\label{eq:gradvelfieldwrong}
    \vec{v} \overset{?}{=} \vec{\nabla}S
\end{equation}
where the $\overset{?}{=}$ symbol indicates that the equality should not be taken at face value.
Since the vorticity is $ \omega = \vec{\nabla}\times\vec{v}, $ note that by applying $\vec{\nabla}\times$ on both sides of~\eqref{eq:gradvelfieldwrong} it would imply that vorticity is never present.
As it is pointed out in detail in Ref~\cite{santosHydrodynamicsVorticesBoseEinstein2016}, the error in~\eqref{eq:gradvelfieldwrong} is a consequence of the application of the chain rule in~\eqref{eq:velfield} without careful consideration of the multivalued nature of $S$.

The apparent consequences of this oversight are specially relevant for the study of turbulence since an immediate result is that the incompressible part of the velocity field also vanishes.
That in turn implies that the incompressible energy spectrum vanishes which would contradict the well-known fact that vortices do exist in BECs.
This contradiction can be corrected by following the reasoning presented by~\citet{santosHydrodynamicsVorticesBoseEinstein2016}.
The velocity field can be corrected by writing
\begin{equation}
    \vec{v} = \vec{\nabla}S + \vec{A}
\end{equation}
where $\vec{A}$ compensates for the discontinuities in $S$ which occurs whenever vortices are present in the system.
According to equation (30) of  reference~\cite{santosHydrodynamicsVorticesBoseEinstein2016} for a general system with $0
\leq S < 2\pi$ the correction must be given by $ \vec{A} = 2\pi\Theta{(R)}\vec{\nabla}\Theta{(I)} $ with $R$ and $I$ being the real and imaginary parts of the wave function respectively, and $\Theta(\ldots)$ the Heaviside step function.
Note that now all the information about the incompressible part of the velocity field, $\vec{v}$, is contained in $\vec{A}$ so that $\vec{v}^{\perp} = \vec{A}^{\perp}$ and $\vec{\nabla}\times\vec{v} = \vec{\nabla}\times\vec{A}$.

One of the main difficulties of directly calculating the incompressible energy spectrum is the particle density in the denominator of the velocity field in Eq.~\eqref{eq:velfield}.
To eliminate this obstacle we shall choose an auxiliary wavefunction, $ \phi \equiv \frac{\sqrt{\pi}}{2}\left(\sign{R} + i~\sign{I}\right) $, here $\sign{\ldots}$ is the sign function.
Such a wavefunction has uniform density and we will demonstrate that it has the same incompressible velocity field and also ends up having some of the same properties of the original turbulent wavefunction, that is, random and uniformly distributed phase, decorrelation between distinct wave numbers and induced wave spectrum in the form of power-laws.

\subsection{Auxiliary Wavefunction}

The auxiliary wavefunction~\eqref{eq:phifield} is defined so that it preserves the incompressible velocity field, but in order
to analytically calculate the energy spectrum we need to verify its statistical properties.
%Particularly we must make sure of which, if any, of the WWT results are still valid for the auxiliary wavefunction.
We are specially concerned if some of the statistical properties of the original wavefunction are present in the auxiliary wavefunction.
\begin{equation}
    \phi \equiv \frac{\sqrt{\pi}}{2}\left(\sign{R} + i~\sign{I}\right)
    \label{eq:phifield}
\end{equation}

First let us consider our system to be in the wave turbulent regime so that $ \tilde{\psi}(\vec{k}) \sim |\vec{k}|^{-\frac{\alpha}{2}}e^{iS} $, with $\alpha \in \mathcal{R}$ being the power-law coefficient of a stationary $n(k) \sim k^{\alpha}$ and $0 \leq S < 2\pi$ a random phase uniformly distributed in this interval.
This stationary state is analytically obtained from the well established Weak Wave Turbulence theory~\cite{proment_sustained_2012, nazarenkoWaveTurbulence2011}.
%which the auxiliary wavefunction will be constructed.%

Numerically, we initialize $\tilde{\psi}(k)$ in a $1024^{3}$ grid in Fourier space, and generate the phase field $S$ as a uniform distribution in the $0 \leq S < 2\pi$ interval for each point $\vec{k}$ in the grid.
We then numerically obtain $\psi(r)$ by using the inverse Fourier transform and construct the auxiliary wavefunction $\phi(r)$ as described by~\eqref{eq:phifield}.
We proceed by applying the direct Fourier transform to obtain $\tilde{\phi}(k)$.

%Numerically, we construct the auxiliary wave function by initializing $\psi$ in a $1024^{3}$ grid, with $\vec{k} \in [-\frac{\pi}{2}, \frac{\pi}{2}]$ and generate $S$ as an uniform distribution in the $0 \leq S < 2\pi$ interval.
%Then we take the inverse Fourier transform to obtain $\psi$ in real space and take $ \phi \equiv \frac{\sqrt{\pi}}{2}\left(\sign{R} + i~\sign{I}\right) $.
%We then take the direct Fourier transform of $\phi$ to obtain $\phi(k)$.

\begin{figure}[ht!]
    \centering
    \includegraphics[width=8.6cm]{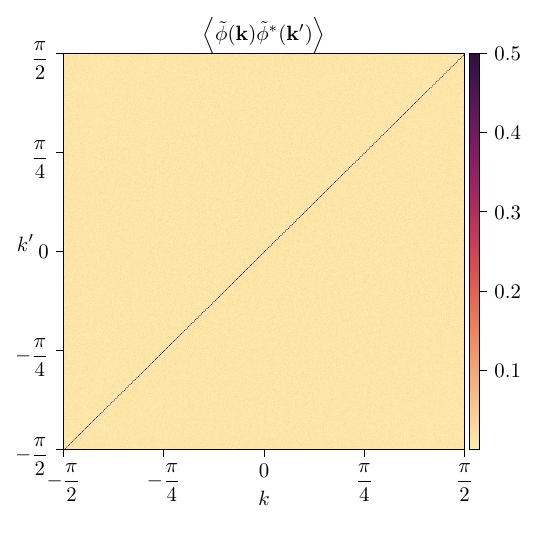}
    \caption{\label{fig:cross}  cross-section of the 2 point correlator of the auxiliary wavefunction $\phi$. Note that the correlation is $0$ everywhere except when $\vec{k} = \vec{k'}$ demonstrating the closure relationship of the wave spectrum, i.e., $\langle{\tilde{\phi}(\vec{k})\tilde{\phi}^{*}(\vec{k'})}\rangle = \langle{|\tilde{\phi}(\vec{k})|^{2}}\rangle\delta(\vec{k} -\vec{k'})$.}
\end{figure}

\begin{figure}[ht!]
    \centering
    \includegraphics[width=8.6cm]{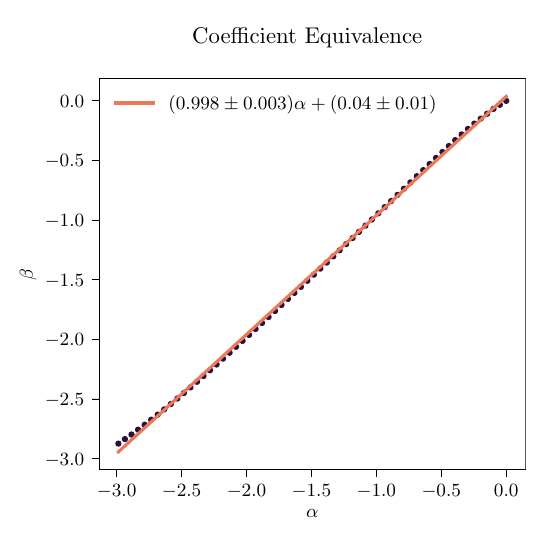}
    \caption{\label{fig:fitalfabeta}  Comparison between power-law coefficients of $n_{\psi}{(k)}$ and $n_{\phi}{(k)}$, the horizontal axis represents the prepared coefficients for the $\psi$ wavefunction and the vertical axis the $\beta$ coefficient obtained from the calculated wave spectrum of the $\phi$ wavefunction. The straight line shows the best fit approximation using a standard linear regression algorithm.}
\end{figure}

We verified that this choice of $\phi$ guarantees that the density of $\phi$ is uniform in the real space and, it turns out, that the phase field also remains random in momentum space.
In addition to the randomness preservation we also see that the closure relationship of the two-point correlation is also preserved for the auxiliary wavefunction as can be seen from, Fig.~\ref{fig:cross} which shows that $\phi$ has no correlation between different wave numbers, this was done by directly calculating $\langle{\tilde{\phi}(\vec{k})\tilde{\phi}^{*}(\vec{k'})}\rangle$, over several runs and averaging the results.
We also observed that $\tilde{\phi}(k)$ will have a power-law wave spectrum, $k^{\beta}$, preserving not only the statistical properties but the nature of the stationary solutions of the kinetic equation.
To obtain the relationship between the coefficients $\beta$ and $\alpha$ we prepared several configurations of $\tilde{\psi}(k)$ with different coefficients, $\alpha$, and calculated the induced power-law coefficient $\beta$ for the auxiliary wavefunction, the results can be seen on Fig.~\ref{fig:fitalfabeta}.
Surprisingly we found that $\beta = \alpha$, in the domain of coefficients predicted by WWT, $-3 \geq \alpha \leq -\frac{7}{3}$, as shown by the best fit coefficients.
Therefore, in the following we will consider $n_{\psi}{(k)} \sim n_{\phi}{(k)}$.

With that we numerically demonstrated that inside the limits of the WWT the auxiliary wavefunction $\tilde{\phi}(k)$ has similar statistical properties to the system's wavefunction $\tilde{\psi}(k)$.
Particularly the wave spectrum of both fields obey the same power-law, this will allow us to analytically calculate the incompressible energy spectrum of the system.
Note that we did all the numerical calculations described by assuming that the system is already in the stationary state predicted by the WWT theory, the existence of which was successfully observed both numerically and experimentally in previously mentioned works~\cite{proment_sustained_2012,hennEmergenceTurbulenceOscillating2009,thompsonEvidencePowerLaw2014,navon_emergence_2016,navonSyntheticDissipationCascade2018}.

\subsection{Incompressible Velocity Field}

A direct calculation of the velocity field, shows that $ \vec{v}_{\phi} = \frac{2}{\pi}\vec{j}_{\phi} $,
with:
\begin{equation}\label{eq:currentphifirst}
    \vec{j}_{\phi} = \frac{\phi^{*}\vec{\nabla}{\phi} - \phi\vec{\nabla}{\phi^{*}}}{2i} = \frac{\pi}{2}\left(\sign{R} + 1\right)\vec{\nabla}\sign{I} - \frac{\pi}{4}\vec{\nabla}[(\sign{R} - 2)\sign{I}]
\end{equation}
where the subscript $\phi$ indicates that the quantity is calculated in relation to the auxiliary wavefunction $\phi$.
This shows that for the auxiliary wavefunction the velocity and current fields are proportional to each other.

Now, taking $ \frac{\pi}{2}\left(\sign{R} + 1\right)\vec{\nabla}\sign{I}, $ from the right-hand side of~\eqref{eq:currentphifirst} and making the following substitution:
\begin{equation*}
    \begin{aligned}
        \sign{a} &= 2\Theta{(a)} - 1,\\
        \Rightarrow \frac{\pi}{2}\left(\sign{R} + 1\right)\vec{\nabla}\sign{I} &= 2\pi\Theta(R)\vec{\nabla}(\Theta(I)) = \vec{A},
    \end{aligned}
\end{equation*}
then:
\begin{equation}\label{eq:currentphilast}
    \vec{j}_{\phi} = \vec{A} - \frac{\pi}{4}\vec{\nabla}[(\sign{R} - 2)\sign{I}]
\end{equation}
where $\vec{A}$ is the vector that carries the information about the incompressible part of the system.
This result shows that $\vec{v}^{\perp} = \frac{\pi}{2}\vec{v}^{\perp}_{\phi}$, that is, the incompressible part of the velocity field of the wavefunction $\psi$ is proportional to that of the auxiliary wavefunction $\phi$.

\subsection{Incompressible Energy Spectrum}

Now we will show that one can infer the statistics of the incompressible velocity field directly from the statistics of the wavefunction itself as long as the system is in a regime where the full statistical description of the waves are known, such is the case of the weak wave turbulent regime.

For an isotropic system it can be shown that:
\begin{equation}
\langle{\vec{v}^{\perp}(\vec{r}){\vec{v}^{\perp*}}(\vec{r}')}\rangle  = \int{\langle{|\tilde{\vec{v}}^{\perp}(\vec{k})|^{2}}\rangle}e^{\vec{k}\cdot{(\vec{r} - \vec{r'})}}d^{3}k, \end{equation}
then, the incompressible kinetic energy per unit mass of the fluid is given by:
\begin{equation}
    E =
    \frac{1}{2}\langle{\vec{v}^{\perp}(\vec{r}){\vec{v}^{\perp*}}(\vec{r})}\rangle  =
    \frac{1}{2}\int{\langle{|\tilde{\vec{v}}^{\perp}(\vec{k})|^{2}}\rangle}d^{3}k =
    \int{\mathcal{E}(k)}d^{3}k.
    \label{eq:kineticenergyperp}
\end{equation}
In order to obtain the spectrum, $\mathcal{E}_{k}$, we need to evaluate the quantity, $ \langle{|\tilde{\vec{v}}^{\perp}(\vec{k})|^{2}}\rangle, $ and as we previously mentioned this is quite difficult to do in general due to the nature of the velocity field.
But as we shall demonstrate if we use the fact that $\vec{v}^{\perp} = \frac{\pi}{2}\vec{v}^{\perp}_{\phi}$, then it becomes straightforward.

From Eq.~\eqref{eq:kineticenergyperp} we can start evaluating the spectrum by making use of the auxiliary wavefunction.
We can obtain the incompressible part of a vector field in $k-\text{space}$ in a general way by using the following relationship
\begin{equation}
    \vec{B}^{\perp} = (\mathbb{I} - \frac{\vec{k}\vec{k}}{|\vec{k}|^{2}})\cdot\vec{B},
\end{equation}
with $\mathbb{I}$ being the identity tensor and $\vec{B}$ a generic vector field.
We also note that
\begin{equation}
    \vec{v}^{\perp} = \frac{\pi}{2}\vec{v}^{\perp}_{\phi} = \vec{j}^{\perp}_{\phi}
    \implies
    \mathcal{E}(k) =
    \frac{1}{2}\left\langle{\tilde{\vec{j}}^{\perp}_{\tilde{\phi}}(\vec{k})\tilde{\vec{j}}^{\perp*}_{\tilde{\phi}}(\vec{k'})}\right\rangle\delta\left(\vec{k} - \vec{k'}\right),
    \label{eq:3Dspectrumphi}
\end{equation}
so calculating the correlation of the particle current of the auxiliary wavefunction is mathematically identical to calculating the same quantity for the velocity field of the original wavefunction.
In order to evaluate~\eqref{eq:3Dspectrumphi} we start by writing $\tilde{\vec{j}}_{\phi}$ as a function of the Fourier transform of the auxiliary wavefunction $\phi$,
\begin{equation}
    \tilde{\vec{j}}_{\phi}(\vec{k}) =
    \frac{1}{{(2\pi)}^{3}}\int{\left(2\vec{k}_{1} - \vec{k}\right)
    \tilde{\phi}(\vec{k}_{1})\tilde{\phi}^{*}(\vec{k}_{1} - \vec{k})}dk_{1}^{3},
\end{equation}
which is done by replacing directly the Fourier representation of $\phi$ into the definition of the particle current, then
\begin{equation}
    \begin{aligned}
        \tilde{\vec{j}}^{\perp}_{\phi}(\vec{k}) & =
        (\mathbb{I} - \frac{\vec{k}\vec{k}}{|k|^{2}})\cdot\tilde{\vec{j}}_{\phi}(\vec{k}) \\
        & =
        \frac{1}{{(2\pi)}^{3}}\int{\left[2\vec{k}_{1} - 2\left(\frac{\vec{k}\cdot\vec{k}_{1}}{|\vec{k}|^{2}}\right)\vec{k}\right]
        \tilde{\phi}(\vec{k}_{1})\tilde{\phi}^{*}(\vec{k}_{1} - \vec{k})}dk_{1}^{3},
    \end{aligned}
\end{equation}
gives us the incompressible part of the particle current which in turn leads to
\begin{equation}
    \begin{aligned}
        \mathcal{E}_{\phi}(\vec{k})
         & = \frac{1}{2}\left\langle{|\tilde{\vec{j}}^{\perp}_{\phi}(\vec{k})|^{2}}\right\rangle\delta(\vec{k} - \vec{k'}) \\
         & =
         \frac{\delta(\vec{k} - \vec{k'})}{{2(2\pi)}^{6}}\int{dk_{1}^{3}}|\vec{M}(\vec{k}, \vec{k}_{1})|^{2}n(\vec{k}_{1})n(\vec{k}_{1} - \vec{k}),
    \end{aligned}\label{eq:currentcorrel}
\end{equation}
where
\begin{equation}
    \vec{M}(\vec{u}, \vec{w}) = 2\vec{w} - 2\left(\frac{\vec{u}\cdot\vec{w}}{|\vec{u}|^{2}}\right)\vec{u},
\end{equation}
and we used the WWT to solve the auxiliary wavefunction correlators in terms of $n(\vec{k})$.

Since $n(\vec{k}) = \mathcal{A}_{k}k^{\alpha}$, where $\mathcal{A}_{k}$ is a proportionality constant dependent on the system initial conditions, we can calculate the right-hand side of~\eqref{eq:currentcorrel} by using the isotropy and homogeneity of the system and the change of variables $\vec{k}_{1} \rightarrow \vec{k}x$ then:
\begin{equation}
    \mathcal{E}_{\phi}(\vec{k}) =\frac{\delta(\vec{k} - \vec{k'})}{2{(2\pi)}^{5}} \mathcal{A}_{k}^{2}k^{2\alpha+5} \int\limits_{{K_{-}}/{k}}^{{K_{+}}/{k}}W(x)dx
\end{equation}
\begin{equation}
    W(x) = x^{4+\alpha}\int\limits_{0}^{\pi}{{(x^{2} + 2x\cos{\theta} + 1)}^{\frac{\alpha}{2}}\sin^{3}(\theta)}{d\theta}
\end{equation}
The limits $K_{\pm}$ represent the boundary of the inertial range in which the cascading solutions, from the WWT, are valid.
Then one can obtain the asymptotic behaviour by expanding $W(x)$ in the limits $k \ll K_{+}$ and $k \gg K_{-}$ and solving the integral up to leading order in $k$.
The asymptotic solutions will be heavily dependent on the value of the power-law coefficient $\alpha$ as can be seen in Eq.~\eqref{eq:3Dspectsol} below:
\begin{equation}
    \mathcal{E}(k) \sim
    \begin{cases}
        k^{\alpha},                    & \text{ if } \alpha < -5,            \\
        k^{-5}\log(k),                 & \text{ if } \alpha = -5, \\
        k^{2\alpha+5} ,                & \text{ if } -5 < \alpha < -\frac{5}{2},  \\
        \log\left(\frac{1}{k}\right) , & \text{ if } \alpha = -\frac{5}{2},  \\
        k^{0} ,                        & \text{ if } \alpha > -\frac{5}{2}.
    \end{cases}\label{eq:3Dspectsol}
\end{equation}
In the literature it is typical to discuss the energy spectrum in terms of the $1D$ spectrum given as $\mathcal{E}^{(1D)}(k) = 4\pi{}k^{2}\mathcal{E}(k)$.
This is done by considering the isotropy of the system and integrating out the solid angle in the last term of Eq.~\eqref{eq:kineticenergyperp}:
\begin{equation*}
    \int{\mathcal{E}(\vec{k})}d^{3}k = \int{4\pi{}k^{2}\mathcal{E}(k)}dk = \int{\mathcal{E}^{(1D)}(k)}dk.
\end{equation*}
Therefore Eq.~\eqref{eq:3Dspectsol} can be written as
\begin{equation}
    \mathcal{E}^{(1D)}(k) \sim
    \begin{cases}
        k^{\alpha + 2},                    & \text{ if } \alpha < -5,            \\
        k^{-3}\log(k),                 & \text{ if } \alpha = -5, \\
        k^{2\alpha+7} ,                & \text{ if } -5 < \alpha < -\frac{5}{2},  \\
        k^{2}\log\left(\frac{1}{k}\right) , & \text{ if } \alpha = -\frac{5}{2},  \\
        k^{2} ,                        & \text{ if } \alpha > -\frac{5}{2}.
    \end{cases}\label{eq:1Dspectsol}
\end{equation}

\section{Discussion}

The spectra obtained in~\eqref{eq:1Dspectsol} are all highly dependent on the WWT coefficient, $\alpha$, therefore the relevant incompressible energy spectra are the ones obtained from the physically valid values of $\alpha$.
The WWT, in the $3D$ case, offers two cascading solutions for the kinetic equation, a waveaction cascade with $\alpha = -\frac{7}{3}$ and an energy cascade with $\alpha = -3$.
Aside from these two solutions there is also the critical balance conjecture, which considers the system when the interaction and kinetic energy are comparable, such situation was observed by~\citet{proment_sustained_2012} with $n(k) \sim k^{-4}$.
With these values we see that the physically relevant results from Eq.~\eqref{eq:1Dspectsol} are:
\begin{equation}
    \mathcal{E}^{(1D)}(k) \sim
    \begin{cases}
        k^{2\alpha+7} , & \text{ if } -5 < \alpha < -\frac{5}{2},  \\
        k^{2} ,         & \text{ if } \alpha > -\frac{5}{2},
    \end{cases}\label{eq:1Dspectsolph}
\end{equation}
The predicted WWT spectra $\mathcal{E}^{(1D)}_{WT} \sim k^{\alpha+4}$ also coincides with the one obtained by our method $\mathcal{E}^{(1D)}(k) \sim k^{2\alpha + 7}$ for the energy cascade solution, which seems to be coincidental since the WWT energy spectra prediction represents the sum of compressible plus incompressible spectra.

\begin{figure}[H]
    \centering
    \includegraphics[width=8.6cm]{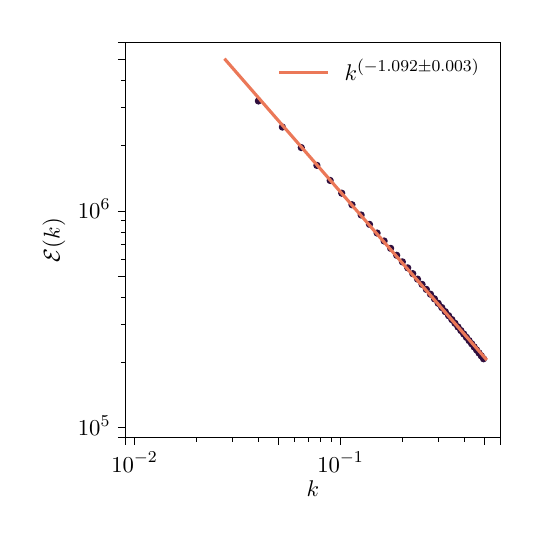}
    \caption{\label{fig:averaged_velocity_energy}  Incompressible kinetic energy spectrum, calculated for $\psi \sim k^{-\frac{3}{2}}$, and best fit (straight line) approximation.}
\end{figure}

\begin{figure}[H]
    \centering
    \includegraphics[width=8.6cm]{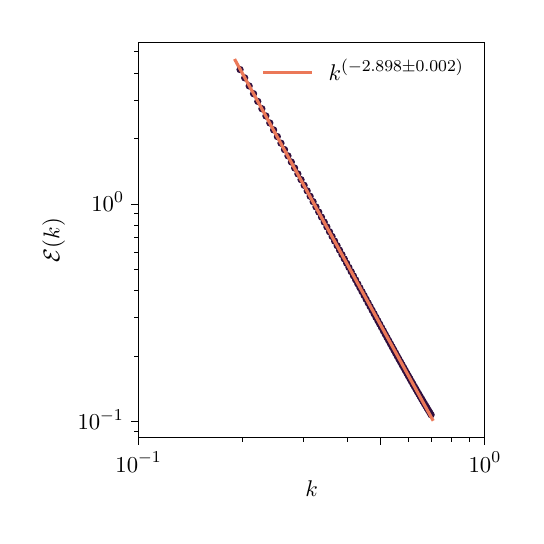}
    \caption{\label{fig:averaged_velocity_critical}  Incompressible kinetic energy spectrum, calculated for $\psi \sim k^{-2}$, and best fit (straight line) approximation.}
\end{figure}

The results predicted by the use of the auxiliary wavefunction can be verified by numerically averaging the velocity field correlator and calculating the spectrum in Fourier representation.
Figures~\ref{fig:averaged_velocity_energy} and~\ref{fig:averaged_velocity_critical} shows the incompressible energy spectrum for $\psi \sim k^{-\frac{3}{2}}$ and $\psi \sim k^{-2}$, respectively.
A linear regression showing the best fit approximation is included in both figures.
The numerically obtained coefficients shows agreement with the ones extracted from~\eqref{eq:1Dspectsolph} for $\alpha = -3$ and $\alpha = -4$.

The spectrum of Fig.~\ref{fig:averaged_velocity_critical}, $\mathcal{E}^{(1D)}(k) \sim k^{-0.898}$, obtained for $\alpha = -4$ is related to the wave spectrum of the critical balance conjecture and is very similar to the incompressible energy spectrum observed in \citet[see Fig.~21]{proment_sustained_2012} near the critical balance regime.
We showed that this spectrum comes from the velocity field statistics as a consequence of the wave spectrum and, since it happens in the transition from the 4-wave to 3-wave regime, it could be indicative of the manifestation of Vinen type turbulence in the transition process, but more investigation would be needed particularly in the vortex decay near the 4-wave to 3-wave transition.

It should be noted that the plots in Fig.~\ref{fig:averaged_velocity_energy} and Fig.~\ref{fig:averaged_velocity_critical} were obtained by preparing a wave function with the characteristics of a WWT solution, in this case the energy cascade and critical balance conjecture, respectively.
After preparing the wavefunction we calculated the velocity field and used the Helmholtz decomposition in order to obtain its incompressible part and calculated the autocorrelation, a multiple sample average was used to reduce the noise from the random phase field.
To finally obtain the spectrum we used a radial average to integrate out the solid angle, followed by the best fit analysis to numerically calculate the angular coefficient.

In summary, we demonstrated that there is a direct relationship between the Kolmogorov--Zakharov power law cascades from wave turbulence and the statistics of the incompressible velocity field of an atomic BEC.
That was done by analytically calculating an analogue to the classical Kolmogorov spectra directly from the statistical distribution of the velocity field, which was possible by using an auxiliary wavefunction that evades the discontinuities of the velocity field and reproduces the statistical properties of the Gross--Pitaevskii wavefunction.
The result obtained not only demonstrates a link between wave turbulence and hydrodynamic turbulence but presentes, as far as the authors know, the first analytical calculation of the velocity field distribution spectrum in atomic BEC.
This is also consistent with the behaviour observed in the literature as both in the numerical simulations done by~\citet{proment_sustained_2012} and the experiments done by~\cite{navon_emergence_2016} it was noted that vortices were present and seemed to play a role in the wave interactions.

Funding: This study was financed in part by CAPES (Coordenação de Aperfeiçoamento de Pessoal de Nível Superior) -- Brasil (CAPES) -- Finance Code 001 and CNPq.
F. E. A. dos Santos also thanks CNPq (Conselho Nacional de Desenvolvimento Científico e Tecnológico, National Council for Scientific and Technological Development) for support through Bolsa de produtividade em Pesquisa (Research productivity scholarship) Grant No. 305586/2017-3 and DAAD-CAPES PROBRAL Grant number 88887.627948/2021-00.

%apsrev4-2.bst 2019-01-14 (MD) hand-edited version of apsrev4-1.bst
%Control: key (0)
%Control: author (72) initials jnrlst
%Control: editor formatted (1) identically to author
%Control: production of article title (-1) disabled
%Control: page (0) single
%Control: year (1) truncated
%Control: production of eprint (0) enabled
%

%\bibliography{emergent_HT_WT_bibliography.bib}

\end{document}